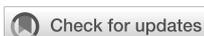

# Dosimetric characterization of single- and dual-port temporary tissue expanders for postmastectomy radiotherapy using Monte Carlo methods

Jose Ramos-Méndez, Catherine Park and Manju Sharma*

Department of Radiation Oncology, University of California, San Francisco, San Francisco, CA, United States

**Purpose:** The aim of this work was two-fold: a) to assess two treatment planning strategies for accounting CT artifacts introduced by temporary tissue-expanders (TTEs); b) to evaluate the dosimetric impact of two commercially available and one novel TTE.

**Methods:** The CT artifacts were managed using two strategies. 1) Identifying the metal in the RayStation treatment planning software (TPS) using image window-level adjustments, delineate a contour enclosing the artifact, and setting the density of the surrounding voxels to unity (RS1). 2) Registering a geometry template with dimensions and materials from the TTEs (RS2). Both strategies were compared for DermaSpan, AlloX2, and AlloX2-Pro TTEs using Collapsed Cone Convolution (CCC) in RayStation TPS, Monte Carlo simulations (MC) using TOPAS, and film measurements. Wax slab phantoms with metallic ports and breast phantoms with TTEs balloons were made and irradiated with a 6 MV AP beam and partial arc, respectively. Dose values along the AP direction calculated with CCC (RS2) and TOPAS (RS1 and RS2) were compared with film measurements. The impact in dose distributions was evaluated with RS2 by comparing TOPAS simulations with and without the metal port.

**Results:** For the wax slab phantoms, the dose differences between RS1 and RS2 were 0.5% for DermaSpan and AlloX2 but 3% for AlloX2-Pro. From TOPAS simulations of RS2, the impact in dose distributions caused by the magnet attenuation was (6.4 ± 0.4) %, (4.9 ± 0.7)%, and (2.0 ± 0.9)% for DermaSpan, AlloX2, and AlloX2-Pro, respectively. With breast phantoms, maximum differences in DVH parameters between RS1 and RS2 were as follows. For AlloX2 at the posterior region: (2.1 ± 1.0)%, (1.9 ± 1.0)% and (1.4 ± 1.0)% for D1, D10, and average dose, respectively. For AlloX2-Pro at the anterior region (-1.0 ± 1.0)%, (-0.6 ± 1.0)% and (-0.6 ± 1.0)% for D1, D10 and average dose, respectively. The impact in D10 caused by the magnet was at most (5.5 ± 1.0)% and (-0.8 ± 1.0)% for AlloX2 and AlloX2-Pro, respectively.

**Conclusion:** Two strategies for accounting for CT artifacts from three breast TTEs were assessed using CCC, MC, and film measurements. This study showed that the highest differences with respect to measurements occurred with RS1 and can be mitigated if a template with the actual port geometry and materials is used.

KEYWORDS

PMRT, temporary-tissue-expanders, Monte Carlo-TOPAS, high-density metal artifacts, collapsed cone convolution algorithm, breast cancer, radiation effects

# 1 Introduction

Post-mastectomy radiation treatment (PMRT) is selectively recommended for patients with locally advanced and/or high-risk biologically aggressive breast cancers (1). For patients who undergo prosthetic breast reconstruction, radiation increases the risk for adverse effects including capsular contracture, scarring at the implant-tissue junction, development of the seroma and dehiscence of the skin incision (2). As such, a two-stage reconstruction using a temporary tissue expander (TTE), followed by PMRT then delayed final prosthetic reconstruction is often preferred (3). The TTEs help preserve the breast skin and organ at risk contours improving the radiotherapy treatment planning, which in turn alleviates the complication risks. Most TTEs consist of an injection port through which a saline solution is injected to expand the surrounding skin. The port consists of a central high-density magnet enclosed in an encasing to locate the injection site (4). In addition, suction drains are routinely placed to drain the seroma (5). The different TTEs such as CPX® (Mentor, Irvine, CA, USA), Natrelle® (Allergan Inc., Santa Barbara, CA, USA) and DermaSpan (Sientra, Inc., Santa Barbara, CA, USA) have a single port with a high-density magnetic disk placed in a high-density encasing. More recently, AlloX2 and AlloX2-Pro (Sientra Inc., Santa Barbara, CA, USA) breast TTEs were introduced with a dual port system. One port is used for traditional saline injection, and the second facilitates fluid drainage. This feature of dual ports enables independent management of postoperative seroma and thereby reducing the rate of infection by 7.8% as shown retrospectively for the AlloX2 TTE (6).

The PMRT is delivered in conventional 2Gy per fraction for a total dose of 50Gy over five weeks, or with more modern hypofractionation techniques over 3 weeks. The 3D CT data is used to delineate tumors and organs at risk (OAR), and the electron density information in the Hounsfield units (HU) of the CT data is used in the calculation of dose distributions. The presence of high-density magnets imposes challenges to accurate treatment planning and delivery. Some key challenges are (1) the increased scatter dose at the skin surface may lead to skin and subcutaneous toxicity varying from mild erythema to skin fibrosis or skin dyspigmentation (2). The tissue attenuation can lead to cold spots or under dosage of the planning target volume (3). The presence of an implant or other high-density materials leads to streaking artifacts that impede the accurate delineation of tumors and OARs. In addition, due to a limited value range of HU to electron density tables in standard CT systems, the density values of TTEs are not reconstructed correctly in the CT data (7), calling into question the accuracy of the computed dose distribution models.

The dosimetric impact in PMRT of single metal ports have been examined in several studies (8–11). Results largely depend on the treatment modality. For example, for 3D-CRT using single 6MV and 15 MV photon beams, the dose perturbations are reported between 5 to 30% (9, 11, 12) and 16% (11), respectively. For VMAT, differences below 6% had been reported (12); however, some studies had reported a negligible difference (13, 14).

The dual ports cover a significant amount of the treatment volume and perturb the radiation treatment field with increased scatter dose and tissue attenuation beneath the device. To the best of our knowledge, there is no literature on the dose perturbations caused by PMRT with dual metal ports. Therefore, this characterization study aims at a detailed comparison of the three TTEs: single port DermaSpan, dual port AlloX2, and the novel AlloX2-Pro. We provide a detailed comparison of the three TTEs using flat, breast phantom geometries and six clinical cases. In addition, the dose computed by the collapsed-cone convolution (CCC) algorithm v5.5 in RayStation TPS is compared with TOPAS Monte Carlo Tool calculations and experimental Gafchromic film measurements.

# 2 Methods

## 2.1 TrueBeam phase space verification

Fifty phase space files containing the positions of particles, angular momenta and kinetic energies generated by Monte Carlo simulations of a 6 MV TrueBeam Linac were obtained from MyVarian at www.myvarian.com/montecarlo. The total number of primary histories per phase space was $10^9$ and was generated without any variance reduction technique. The phase spaces were scored at a plane positioned at 73.3 cm from the Linac isocenter, upstream of any moving parts of the Linac treatment head. A comparison was performed between the percentage depth-dose and lateral dose distributions at several depths calculated in water and measured data obtained at the time of commissioning for a TrueBeam Linac at our institution. For that, two open field setups at 3 x 3 cm$^2$ and 10 x 10 cm$^2$ defined at 100 cm SSD were used. The water phantom had dimensions of 20 x 20 x 35 cm$^3$ with a voxel resolution of 1 x 1 x 0.5 mm$^3$; the highest resolution was used along the beam direction. The following linac devices were included in the simulation: jaws, base plate, 120 Millennium MLC, and mylar tray. The geometry details were obtained from the vendor. The absorbed dose averaged by primary history retrieved at 10 cm depth was used to scale the simulations to the dose calibration conditions at our institution: 1 cGy/MU at a depth of maximum dose for a 10 x 10 cm$^2$ field defined at 100 cm SSD. An exponential fit was adjusted to the calculated PDD between the range of 5 to 15 cm to retrieve the calculated absorbed dose at 10 cm depth.

The Monte Carlo simulations were performed with TOPAS version 3.7 (15, 16) built on top of Geant4 toolkit version 10.07 patch 3 (17). The physics list was the electromagnetic module called "g4em-standard_opt4" which was described and benchmarked for its application in radiotherapy as reported elsewhere (18). For all dose calculations, azimuthal particle redistribution with a split number of 50 (19) was used through the geometrical particle split technique available in TOPAS (20). The statistical uncertainty of the dose distributions was 0.5% or better in all simulated cases.

## 2.2 Breast tissue expander geometries

Breast TTEs consisted of a silicon bag filled with saline solution containing one or two draining or filling ports with a high-density magnet embedded to allow its localization. Three breast TTEs were used in this work. Two commercially available (DermaSpan™ and AlloX2®) and a novel TTE (AlloX2-Pro-Sientra, Inc). The geometry details and materials of the ports obtained from Sientra Inc. are presented in Figure 1. The DermaSpan model consisted of a single titanium (ρ=4.54 g/cm$^3$) port with a neodymium (ρ=7.6 g/cm$^3$) magnet enclosed. The AlloX2 model consisted of two titanium ports with one neodymium magnet enclosed in each port. The AlloX2-Pro model consisted of two ports made of peek material (ρ=1.3 g/cm$^3$), with a single neodymium magnet located between the ports. The geometry and densities from all the three ports were saved as contour templates in RayStation.

## 2.3 Strategies for handling metal artifacts

The CT artifacts caused by the metal-ports are managed using two density override strategies at our institution. The first strategy (hereafter called RS1) consists of identifying the metal by adjusting the image window-level to display only the brightest region, assumed occupied by the metal port. Subsequently, a contour is delineated enclosing the artifact and the density of surrounding voxels is set to unity. The second strategy (hereafter called RS2) consists of registering rigidly a geometry template with the dimensions, materials, and densities from the corresponding metal-ports obtained from the vendor; the density of voxels outside the port geometry is set to unity. Both strategies were compared using Collapsed Cone Convolution (CCC) version 5.5 in RayStation version 11A, and TOPAS Monte Carlo simulations. The resolution of the dose grid for RayStation and TOPAS calculations was 2 x 2 x 2 mm$^3$. Calculated results were compared with Gafchromic film (Ashland Inc.) measurements using two irradiation setups as described below.

## 2.4 Wax slab phantom setup

A setup consisting of a wax slab phantom irradiated by an AP field was configured to assist in the validation of TOPAS simulations for each TTE port. For each TTE, the ports were stripped off from the silicon bag and embedded in a slab phantom made of wax (ρ=0.92 g/cm$^3$). The phantom had dimensions of 30 x 30 x 1.7 cm$^3$. An irradiation setup was configured consisting of the slab phantom stacked between 1.5 cm thickness of plastic water and 10 cm thickness underneath, see Figure 2. An iterative metal artifact reduction (iMAR) (Siemens Medical System) algorithm was used to reduce the high-density metal artifacts. The setup was simulated and exported to RayStation TPS for planning. The plan consisted of a 6 MV field of 15 x 10 cm$^2$ defined at 100 SSD, 500 MU delivered in the AP direction. The remained metal artifacts were handled with the two strategies described in section 2.3. The setup was reproduced with TOPAS simulations which included the actual port geometries shown in Figure 1. The ports were aligned to the metal artifact using the RayStation contours from RS2 as a frame of reference. The overlapping of geometries was handled by the feature Layered Mass Geometry (21). Film dosimetry was performed by placing Gafchromic films at different positions as shown in Figure 2.

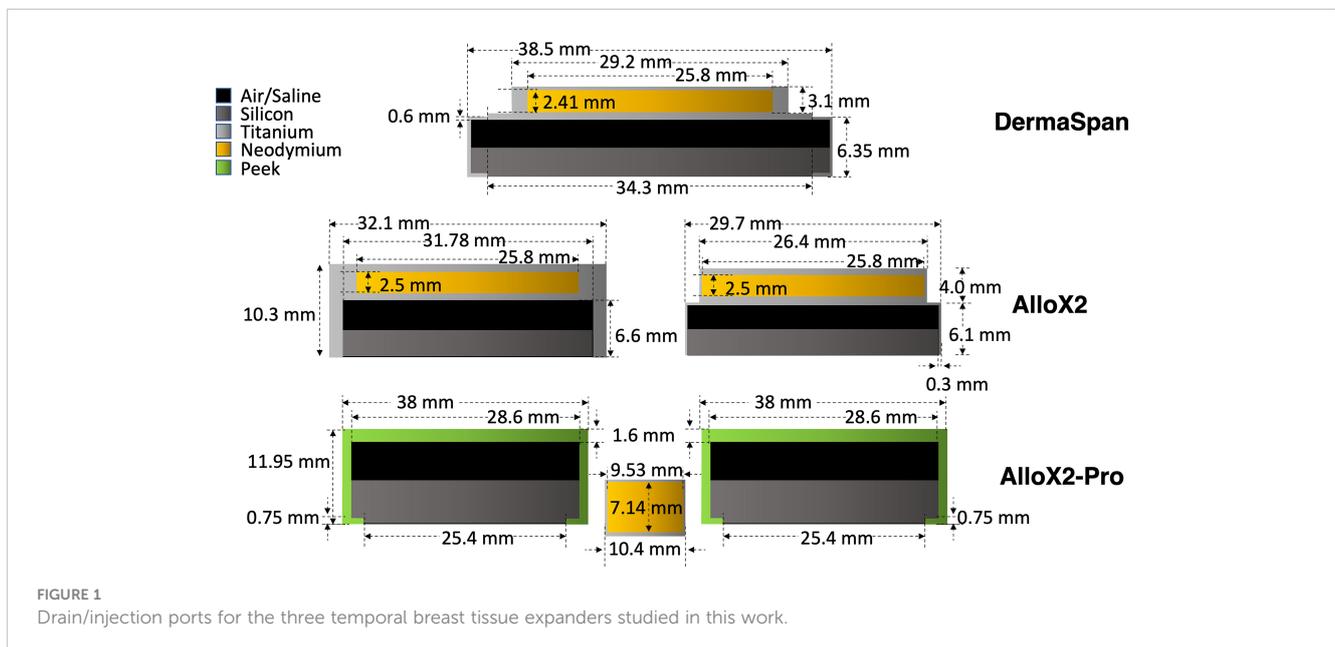

FIGURE 1
Drain/injection ports for the three temporal breast tissue expanders studied in this work.

## 2.5 Breast tissue expander phantom setup

The effect of using multiple gantry angles was evaluated for AlloX2 and AlloX2-Pro TTEs. The partial arc irradiations were performed on the ports using an open field as detailed below. This setup was representative of a worst-case scenario where multiple x-ray beams interacts with the metal port for most of the irradiation time.

The AlloX2 and AlloX2-Pro TTEs were irradiated in their standard configuration during PMRT i.e., embedded in the silicon bag filled with water. The silicon bag wall (~1.1 g/cm$^3$) was about 1 mm of thickness and had a negligible effect on the dose distributions. In this work, water was used instead of saline solution which shown to be dosimetrically equivalent for MV radiation. However, it has a dosimetric impact by 5% for kV photons, as shown by (22). A customized breast phantom holder and bolus (5 mm thickness) were made with wax to immobilize the phantom for reproducibility. The bolus was placed on top of a thermoplastic mesh covering the breast tissue expander with the air gaps filled with superflab bolus as best as possible. CT images were obtained with iMAR algorithm (section 2.3) and exported to RayStation TPS for planning. The plan consisted of a 6 MV conformal arc (3 x 3 cm$^2$), gantry angles from 90 to 270 degrees in the counterclockwise direction, delivering 355 MU in a single fraction, see Figure 2. The partial arc configuration considered the contribution of parallel opposed fields at 90 and 270 deg. Contours were drawn for the analysis which included the silicon bag, an expanded wall to the silicon bag of 3 mm thickness split into four contours. These contours covered the anterior (C_Anterior), posterior (C_Posterior), left (C_Left) and right (C_Right) directions of the beam. Pieces of films were positioned at several depths as shown in Figure 2. The films for analysis were 1 x 1 cm$^2$ and were read at least 24 hours after the irradiation.

# 3 Results

## 3.1 TrueBeam phase space verification

In Figure 3, the measured percentage depth-dose (PDD) and crossline dose profiles are compared with the ones calculated with the Varian phase spaces for two open fields. For the crossline profiles, several curves are displayed at a depth of 1.5 cm, 10 cm, and 20 cm depths. The bottom of each panel displays the γ-index value resulting from the TOPAS and measurements comparison. As

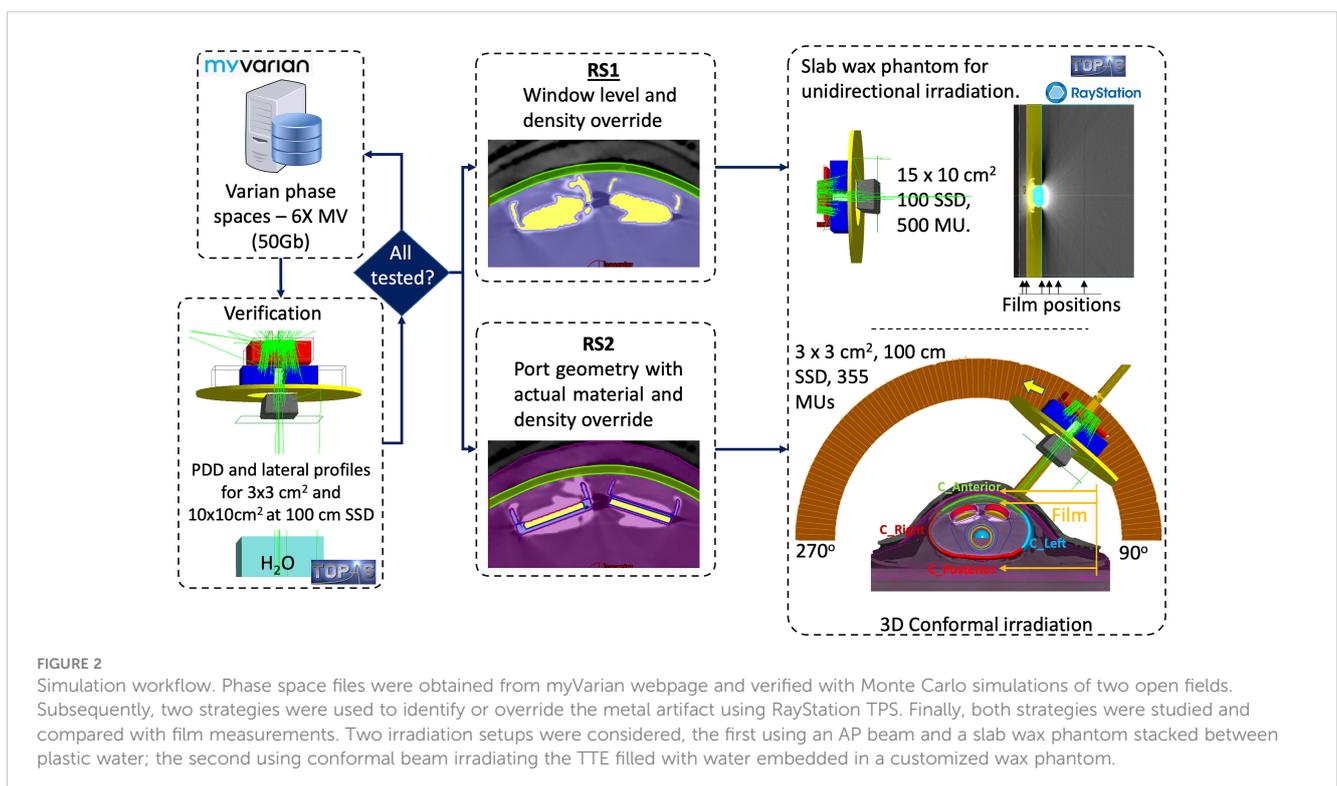

FIGURE 2
Simulation workflow. Phase space files were obtained from myVarian webpage and verified with Monte Carlo simulations of two open fields. Subsequently, two strategies were used to identify or override the metal artifact using RayStation TPS. Finally, both strategies were studied and compared with film measurements. Two irradiation setups were considered, the first using an AP beam and a slab wax phantom stacked between plastic water; the second using conformal beam irradiating the TTE filled with water embedded in a customized wax phantom.

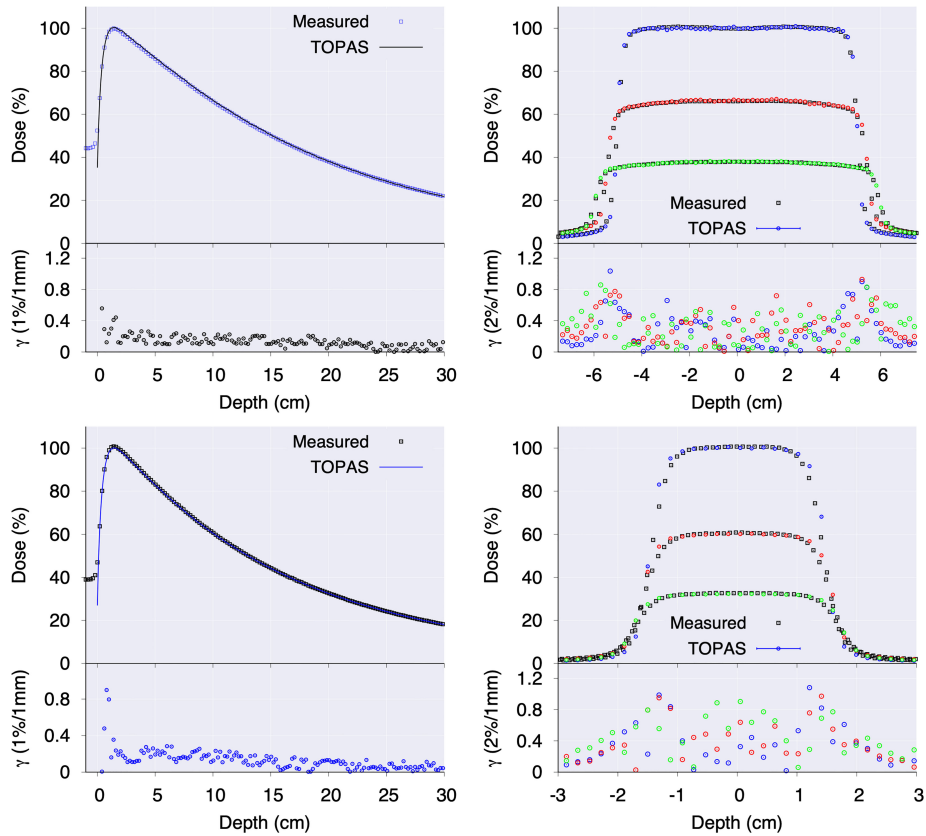

FIGURE 3

Percentage depth-dose and crossline profiles for 10 x 10 cm² (top row panel A and B) and 3 x 3 cm² (bottom row **C**, **D**) fields calculated with Varian phase spaces. Crossline profiles **(B, D)** are presented at depth of maximum dose, and at 10 cm and 20 cm depth. The γ-index values are presented at the bottom of each panel.

depicted, for all the panels the γ-index is below unity for the 1%/1mm (PDD) and 2%/1 mm (crossline) criteria.

## 3.2 Slab wax phantom setup

Panels of Figure 4 show depth-dose profiles for the configuration consisting of the breast tissue expander ports embedded in a wax slab phantom. For DermaSpan and AlloX2-Pro the central profiles are shown, whereas for AlloX2, the profiles crossing the injection port are shown. Film measurements are shown with symbols. At the bottom of the panels, the less restricted of percentage difference and distance-to-agreement to the measured data are shown. The vertical lines delimit the region occupied by the slab wax phantom. For the DermaSpan port (panel A), both RS1 and RS2 were within 2%/1 mm in the buildup region

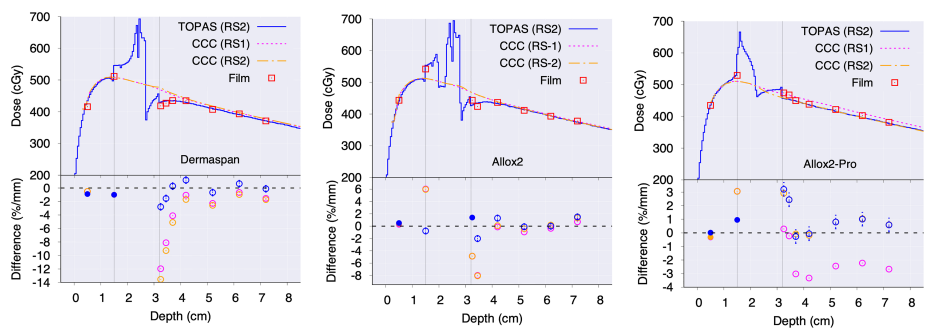

FIGURE 4

Depth-dose curves comparing the CCC (RS1 in dashed, RS2 in dotted-dashed), TOPAS (RS2 in solid) and film measurements (empty squares) for: DermaSpan **(A)**, AlloX2 **(B)** and AlloX2-Pro **(C)**. The bottom of each panel shows the least restricted between the percentage difference (empty symbols) and distance-to-agreement (filled symbols). The vertical lines limit the region occupied by the slab wax phantom.

and distal falloff. Much higher difference was seen at the region immediately at downstream the port. For the AlloX2 (panel B), both RS1 and RS2 were within 2%/1 mm at the buildup and distal falloff. Lastly, for the AlloX2-Pro (panel C), at the buildup region, both RS1 and RS2 were within 2%/1 mm from measurements. At the distal falloff, RS1 differed from the film measurements by 2.7% whereas for RS2 the differences were within 1%. For all the three ports, TOPAS simulations were within (2% ± 0.5%)/1mm in the buildup and distal falloff regions.

The effect of the port in the depth-dose distributions outside the wax phantom was calculated with TOPAS by comparing simulations with the port substituted by wax. Results are shown in the Figure 5. As depicted, the underdosages caused by the attenuation from the magnets in the ports were (6.4 ± 0.4)%, (4.9 ± 0.7)% and (2.0 ± 0.9)% for DermaSpan, AlloX2 and AlloX2-Pro, respectively. In the region proximal to the beam entrance, an overdose caused by the backscatter radiation was observed for the AlloX2-Pro. The overdose decays rapidly from about 3% ± 1% to zero within the first 5 mm.

## 3.3 Tissue expander phantom setup

In panels of Figure 6 the dose profiles along the anterior-posterior direction traversing the drain and central magnets are shown for the AlloX2 (panel A) and AlloX2-Pro (panel B), respectively (section 2.5 and Figure 2). Film measurements are shown with symbols at three positions. For both TTEs, RS2 calculated with CCC (CCC (RS2)) agreed reasonably well with TOPAS calculations but did not reproduce the dose perturbation near the magnet, at about the 4 cm position. RS1 (MC (RS1)) and RS2 (MC (RS2)) results calculated with TOPAS had better agreement to the film measurements, been RS2 the closer to the measured data, as shown in the bottom of each panel of Figure 6. The axial isodose distributions calculated with TOPAS for AlloX2 and AlloX2-Pro using RS-1 and RS-2 are displayed in Figure 7. As depicted, the most significant dose differences, as large as 25% ± 1.5% and 28% ± 1.5%, occur locally around the magnet region. These dose differences are almost entirely contained by the silicon bag. The dosimetric impact outside of the bag is minimal as shown for the contour volumes in Tables 1, 2.

The impact of the TTE port in the dose distribution was quantified by comparing dose volume histogram (DVH) parameters for simulations with and without the metal port, for the contours displayed in Figure 2. Results are shown in Tables 1, 2 for the AlloX2 and AlloX2-Pro, respectively. Combined statistical uncertainties were 1.0%, one standard deviation, or better. For AlloX2, the impact of the metal port calculated by RS1 and RS2 exceeded statistical uncertainties only for the contour C_Posterior located at the posterior region of the phantom, effect caused by the attenuation introduced by the metal port. In this region, RS1 produced a higher dose than using RS2, e.g., by 2.1% for D10. On the other hand, for AlloX2-Pro the impact of the metal port in the computation of DVH parameters shown in Table 2 resulted in subpercentage differences, smaller than the combined statistical uncertainty. Furthermore, RS1 and RS2 were statistically equivalent as the percentage differences between DVH parameters fell within the combined statistical uncertainty.

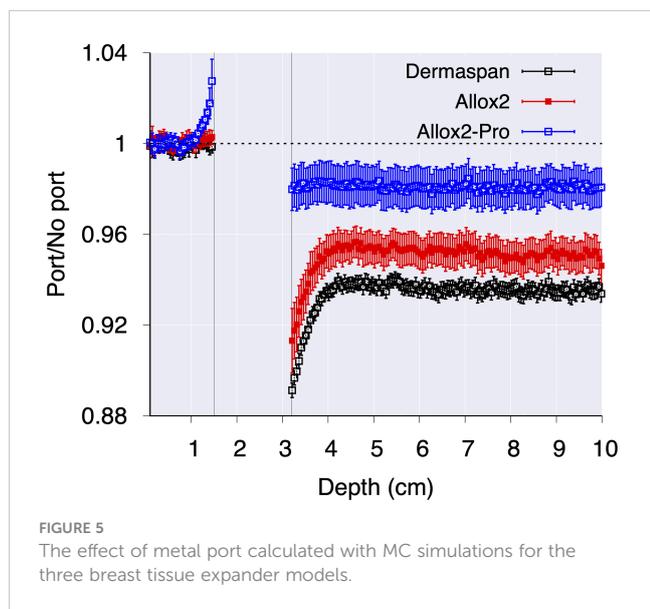

FIGURE 5
The effect of metal port calculated with MC simulations for the three breast tissue expander models.

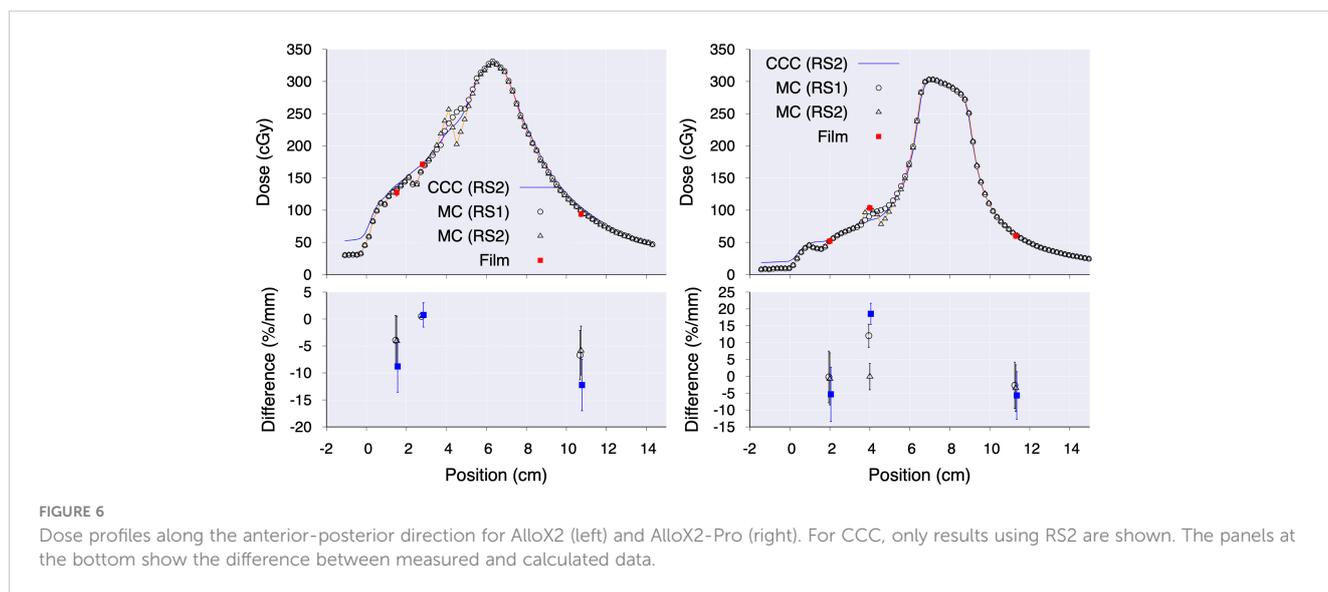

FIGURE 6
Dose profiles along the anterior-posterior direction for AlloX2 (left) and AlloX2-Pro (right). For CCC, only results using RS2 are shown. The panels at the bottom show the difference between measured and calculated data.

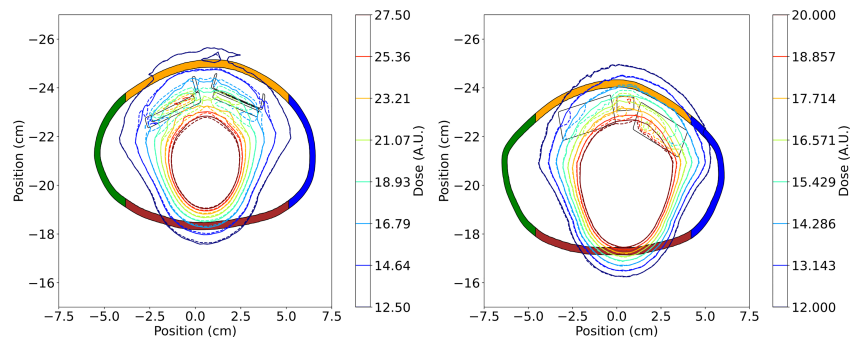

FIGURE 7
Axial isodose distributions in arbitrary units (A.U.) calculated with Monte Carlo for the AlloX2 (left) and AlloX2-Pro (right) TTEs. Solid lines correspond to RS1, and dashed lines correspond to RS2. The colored regions correspond to each contour shown in Table 2. The outer limiting frame of the metal ports are shown with solid lines.

## 4 Discussion

In this work, the dosimetric characterization of three TTEs was performed with the Monte Carlo method and CCC. Dose at selected positions in two irradiation setups, using wax slab phantom (3D-CRT) and customized breast phantom (conformal arc radiotherapy), were compared with film measurements obtaining an overall agreement within 3%. For both irradiation setups, two strategies for handling the CT artifacts produced by TTE metal ports in the calculation of dose distributions for were evaluated.

For the 3D-CRT irradiation setup, the absorbed dose for DermaSpan and AlloX2 was attenuated downstream the magnet. The thickness of each magnet was 2.41 mm and 2.5 mm for DermaSpan and AlloX2, respectively (Figure 1). Under ideal conditions neglecting scattering, the attenuation caused by the magnet (7.4 g/cm$^3$) irradiated with MV x-rays was expected to be ~5% approximately, the Monte Carlo calculated results also included the titanium port and resulted 6.4% and 4.9%, respectively (Figure 5). Conversely, for the AlloX2-Pro (7.14 mm thickness) the attenuation was substantially lower. This effect was caused by the magnet geometry; the physical dimensions perpendicular to the beam were about one third smaller than for the other two ports. Thus, there was more in-scatter radiation from

TABLE 1 Impact of TTE port in dose distributions for AlloX2.

| ROI | Vol. cc | No port – RS1 (%) | | | No port – RS2 (%) | | | RS1–RS2 (%) | | |
|---|---|---|---|---|---|---|---|---|---|---|
| | | D1 | D10 | Ave. | D1 | D10 | Ave. | D1 | D10 | Ave. |
| **C_Left** | 7.9 | 0.1 | -0.1 | 0.4 | 0.0 | -0.2 | 0.7 | 0.0 | -0.1 | 0.3 |
| **C_Right** | 8.3 | -0.1 | -0.1 | 0.2 | -0.1 | 0.0 | 0.4 | 0.0 | 0.1 | 0.2 |
| **C_Anterior** | 17.6 | -0.2 | -0.1 | -0.2 | -0.3 | -0.2 | -0.3 | -0.1 | -0.1 | -0.1 |
| **C_Posterior** | 15.4 | 4.3 | 3.7 | 2.6 | 6.2 | 5.5 | 4.0 | 2.1 | 1.9 | 1.4 |

The impact is quantified by the percentage differences between DVH parameters calculated with Monte Carlo. The percentage difference between RS1 and RS2 are also shown. DVH parameters include the dose at 1% of the volume (D1), dose at 10% of the volume (D10) and average dose.

TABLE 2 Impact of TTE port in dose distributions for AlloX2-Pro.

| ROI | Vol. cc | No port – RS1 (%) | | | No port – RS2 (%) | | | RS1–RS2 (%) | | |
|---|---|---|---|---|---|---|---|---|---|---|
| | | D1 | D10 | Ave. | D1 | D10 | Ave. | D1 | D10 | Ave. |
| **C_Left** | 9.0 | -0.1 | -0.2 | -0.2 | -0.5 | -0.4 | -0.6 | -0.4 | -0.3 | -0.4 |
| **C_Right** | 9.1 | -0.1 | 0.2 | -0.1 | -0.6 | -0.2 | -0.6 | -0.5 | -0.4 | -0.5 |
| **C_Anterior** | 15.8 | -0.1 | -0.2 | -0.2 | -1.0 | -0.9 | -0.7 | -1.0 | -0.6 | -0.6 |
| **C_Posterior** | 18.1 | 0.0 | 0.4 | -0.1 | -0.5 | -0.2 | -0.6 | -0.6 | 0.6 | -0.5 |

The impact is quantified by the percentage differences between DVH parameters calculated with Monte Carlo. The percentage difference between RS1 and RS2 are also shown. DVH parameters include the dose at 1% of the volume (D1), dose at 10% of the volume (D10) and average dose.

the unobstructed portion of the beam for AlloX2-Pro. The in-scatter radiation compensated the attenuation of dose leading to an underdose of about 2%. On the other hand, in the buildup region backscatter dose was observed for AlloX2-Pro only. That backscatter originated by the closer position of the AlloX2-Pro magnet to the TTE surface compared to the other two models. In the literature, backscatter dose factors for 6 MV beams incident in lead (11.3 g/cm$^3$) had been reported to reduce from a factor of 1.03 to 1 within the first centimeter (23). The dose profile calculated with Monte Carlo in this work showed the equivalent behavior as that reported in the literature. The clinical impact of the backscatter dose is expected to be negligible as the maximum extent of the magnet dictates the diameter of the region irradiated by backscatter radiation. This diameter (10.4 mm) is smaller than the diameter stated by ICRU 50 (15 mm) for the definition of a hot spot (24).

The calculated absorbed dose using CCC for both strategies (RS1 and RS2) agreed with Monte Carlo and film measurements within 2%/1 mm for DermaSpan and AlloX2 and within 3%/1 mm for AlloX2-Pro in the buildup region and distal falloff of the depth-dose distribution (Figure 4).

The higher discrepancies occurred in the region downstream of the magnet within the first centimeter. While this discrepancy was a result of the limitations of the dose calculation algorithms and dose grid resolution, its location was expected to be within the filled TTE silicon bag region which might encompass at least 4-5 cm thickness, having minimal impact on the patient. The closer agreement between RS1 and RS2 for DermaSpan and AlloX2 at the distal falloff region was not surprising. The maximum density (2.5 g/cm$^3$) from the CT density tables assigned to the metal artifact for RS1 was about three times smaller than the actual magnet density (7.4 g/cm$^3$) used in RS2, however, the thickness of the identified metal artifact was also about three times greater than the thickness of the actual magnet geometry. Thus, the amount of attenuation in both cases was similar. On the other hand, for AlloX2-Pro the thickness of the metal artifact and the magnet were about the same dimension. Therefore, there was less attenuation using strategy RS1 that led to an overdose of about 3% compared with RS2.

The impact of the beam direction was quantified using partial arc irradiation and a customized phantom for both AlloX2 and AlloX2-Pro. Comparison between simulations with and without port were presented in Tables 1, 2. For AlloX2, the metal port attenuated the dose distribution posteriorly leading to a reduction of the D10 parameter by 5.5%, calculated with RS2. By using RS1, this value can be overestimated by ~2% as shown in Table 1. For the regions located at the lateral positions, the effect of the port was mitigated by the opposed radiation fields (23). This compensation resulted in a negligible difference in the DVH parameters as shown in Table 1. On the other hand, for AlloX2-Pro the impact of the metal port under partial arc irradiation resulted in sub-percentage differences in the DVH parameters, as shown in Table 2. This effect resulted from the small size of the magnet, which allowed more contribution from the in-scatter radiation, as shown for the slab wax phantom setup. Finally, sub-percentage differences in DVH parameters between RS1 and RS2 resulted from the comparable dimension of the metal artifact and the magnet.

In this work 6 MV beams were considered. Retrospective studies reported that 6 MV beams are mostly used for 3D planning of breast with tangents (25) while higher energy beams are often used for large breast separations to improve homogeneity. Above 10 MV, the dose distributions are highly affected by the pair production process within the first 2 cm from the surface of metal objects (23). In addition, the production of photoneutrons takes relevance. Contrary to CCC, these two interaction processes can be explicitly modeled with the Monte Carlo method so that dose differences between the two methods are expected near the metal ports. The dosimetric study of high energy beams in metal ports is out of the scope of current work as a prior validation of TOPAS for the simulation of the photoneutrons yield is needed. This task is ongoing in our research group and will be presented in future work.

In a typical IMRT treatment in VMAT mode, for example, the MLC modulation might partially or totally occlude the radiation directed to the metal port. Thus, the partial arc configuration represented the extreme scenario when the port was irradiated all the time. The highest differences found in the DVH parameters calculated with RS1 and RS2 might be mitigated by the MLC modulation.

Finally, caution must be practiced when higher saturation HU values are used, which lead to higher density values. The density assigned to the identified metal artifact in RS1 highly depended upon the CT density tables and the delineation of artifacts. Thus, we recommend using the actual port geometry and materials. Templates compatible with RayStation TPS are provided in the supplementary material of this work to reduce the delineation time for the TTEs studied in this work.

## 5 Conclusions

The dosimetric impact of the TTEs in PMRT depended on the geometry, artifact delineation method, and irradiation conditions. The greatest differences with respect to measurements were observed in the RS1 strategy. Using a template with the actual port geometry and materials (RS2) can alleviate the differences and reduce the artifact delineation time. Negligible dose perturbation was observed for the novel TTE under continuous partial arc irradiation conditions compared to a single beam at normal incidence.

## Data availability statement

The raw data supporting the conclusions of this article will be made available by the authors, without undue reservation.

## Author contributions

MS was involved in conceptualizing the research design, planning, and continued supervision of the work. MS performed and processed the experimental data and analysis and wrote on the manuscript. JM conceptualized and executed the Monte Carlo simulations, analyzed the data, and wrote the manuscript. CP aided in interpreting the results and worked on the manuscript. All authors contributed to the article and approved the submitted version.

# Funding

This work is supported in part by a sponsored grant sponsored by Sientra Inc CA0160869 and NIH/NCI U24 CA215123.

# Acknowledgments

We acknowledge Peter Li from the University of California Berkely for his help in analyzing the film data, Annette Villa and Naoki Dominguez-Kondo from the University of California San Francisco for making the wax phantoms and for providing the TOPAS geometry extension for the 120 Millennium MLC, respectively; and Darren Sawkey from Varian for providing the 120 Millennium MLC geometry details. We would like to thank Joanna C Yang, Adam Melancon and Junhan Pan for the initial discussions on AlloX2 TTEs.

# Conflict of interest

This work was partially supported by a research grant from Sientra Inc.

# Publisher's note

All claims expressed in this article are solely those of the authors and do not necessarily represent those of their affiliated organizations, or those of the publisher, the editors and the reviewers. Any product that may be evaluated in this article, or claim that may be made by its manufacturer, is not guaranteed or endorsed by the publisher.